\documentclass[12pt]{article}
\usepackage{amssymb}
\usepackage{epsfig}
\def\ostrutdim{3ex} \def\ustrutdim{1ex}
\def\ostrut{\vrule height \ostrutdim width 0pt \relax}
\def\ustrut{\vrule width 0pt depth \ustrutdim \relax}
\def\oustrut{\ostrut\ustrut}  

\def\gev{$\mbox{GeV}^2$}
\def\lqcd{$\Lambda_{\mathrm{QCD}}$}

\begin{document}
\setlength{\baselineskip}{0.75cm}
\setlength{\parskip}{0.45cm}
\begin{titlepage}
\begin{flushright}
\large
DO-TH 95/14\\ August 1995
\end{flushright}
\normalsize
\vspace{1.5cm}
\begin{center}
\LARGE
{\bf BFKL versus HERA}\\
\vspace{2.5cm}
\large
I.\ Bojak and M.\ Ernst\\
\vspace{1cm}
Institut f\"ur Physik, Universit\"at Dortmund \\
D--44221 Dortmund, Germany
\end{center}
\vspace{3.5cm}
\underline{{\large{Abstract}}} \\[2ex]
\normalsize
\setlength{\baselineskip}{0.75cm}
\setlength{\parskip}{0.45cm}
The BFKL equation and the $k_T$-factorization theorem are used to
obtain predictions for $F_2$ in the small Bj{\o}rken-$x$ region over
a wide range of $Q^2$. The dependence on the parameters, especially
on those concerning the infrared region, is discussed.
After a background fit to recent experimental data obtained at HERA and
at Fermilab (E665 experiment), we find that the predicted, almost $Q^2$
independent BFKL slope $\lambda\gtrsim 0.5$ appears to be too steep at
lower $Q^2$ values. Thus there seems to be a chance that future HERA
data can distinguish between pure BFKL and conventional field theoretic
renormalization group approaches.
\end{titlepage}
\newpage
%
%
%
\section{Introduction}
The usual kinematic variables used
for discussing deep-inelastic electron-proton scattering
are derived from the four-momenta $p$ of the incoming proton
and $q$ of the exchanged virtual photon: $Q^2\equiv -q^2$ and
the Bj{\o}rken variable $x\equiv Q^2/2\, p\cdot q$.
In the region where higher twist effects are likely to be negligible, i.\ e.,
for $W^2\equiv Q^2\,(1/x-1)\gtrsim 10$ \gev , the
(Dokshitzer-Gribov-Lipatov-)Altarelli-Parisi [(DGL)AP] set of equations
\cite{ap,dgl} describes the evolution of the structure function $F_2$ with
$Q^2$ very well. In leading order (LO) all powers of
$\alpha_s\,\ln (Q^2/\mu^2)$ are summed by the AP evolution equations,
which take into account just strongly ordered parton-$k_T$ ladders.
Nowadays usually the next-to-leading order (NLO) set is used, where terms
of the form $\alpha_s^n\,\ln^{n-1}(Q^2/\mu^2)$ are also summed,
taking {\em non-\/}ordered $k_T$ contributions (covariantly) into account
as well. Even in NLO, the AP equations do obviously not contain all
leading logarithms in $x$. Thus one might na\"\i vely expect
the AP framework to break down at some small value of $x$, where a
resummation of all powers of $\alpha_s\,\ln(1/x)$ should be necessary,
although no perturbative instability between LO and NLO has been
observed thus far in the presently relevant kinematic regime
\cite{grv92,grs,grv94}.

Such a resummation in LO is provided by the Balitskij-Fadin-Kuraev-Lipatov
(BFKL) equation \cite{bfkl}. The equation treats the gluons only,
which are expected to be the dominant partons at small $x$.
This might be deduced from the AP splitting functions \cite{ap},
which become $\sim 1/x$ for splitting to gluons and constant for splitting to
quarks in the small $x$ limit. It should be remarked though,
that this argument may be too simple, as it neglects the influence of the
particular shape of the parton distributions used \cite{grs,grv94} and
furthermore does not respect the fundamental energy-momentum conservation
constraint.
The BFKL resummation is formally correct for all $Q^2$, but
a fixed coupling $\overline{\alpha}_s$ was used in its derivation.
Furthermore it is based on LO perturbative QCD, using (non-covariant) $k_T$
cut-off regularizations, thus we are limited to
sufficiently large $Q^2$. Since the resummation assumes that subleading
terms in $x$ are small, including those involving logarithms of $Q^2$,
it will also not be valid at high $Q^2$.
A conservative range for $Q^2$ would be 4 to 50 \gev , and the range
0.8 to 120 \gev\ explored in this work should be regarded as the
extreme limit.

To estimate the effects of the subleading terms, a NLO resummation
in $x$ would be necessary. But even though there has been some progress
in that direction \cite{xnlo}, a final result has not yet been obtained.
Ultimatively it should even be possible to find a unified evolution equation
covering the whole perturbative region \cite{unify,ColEl,KMS}.
But a calculation
that can be confronted with experiment is still missing,
thus we stay with the usual BFKL formalism for the time being.

Since a fixed coupling constant seems unreasonable in view of the
running coupling of the AP framework we wish to connect to, the replacement
$\overline{\alpha}_s\rightarrow\alpha_s(Q^2)$ is done by hand.
There is really no rigorous motivation for this step. Some trust
in this procedure can be gained by considering the representation
of the evolution equations as ladder diagrams. It is well known that
the LO AP equations can be represented in a physical gauge by a sum of
ladder graphs which are strongly ordered in the transverse momentum
$k^2$: $Q^2\gg k^2\gg k_{n-1}^2\gg\ldots\gg k_1^2\gg k_0^2$.
In the small $x$ limit of the AP equation we consider only the dominant
gluon ladder (see Fig. \ref{leiter}) and keep only the terms
with double leading logarithms (DLL)
$\alpha_s\,\ln (1/x)\,\ln(Q^2/\mu^2)$. This corresponds to introducing
an additional strong ordering in $x$: $x\ll x_{n-1}\ll\ldots\ll x_0$.
The BFKL evolution can also be described in terms of a (reggeized) gluon
ladder, using the strong ordering in $x$ only, to get the
leading logarithms in $x$. If we let the coupling run, then
BFKL will reduce to LO DLL upon imposing the ordering in transverse momentum.

The main feature of the BFKL evolution with fixed coupling constant
$\overline{\alpha}_s$ is the growth of the unintegrated gluon
$\sim x^{-\lambda}$ with $\lambda=\frac{3\,\overline{\alpha}_s}{\pi}\,4\,\ln 2
\simeq 0.5$.
Due to the dominance of the gluons at small $x$ we expect a corresponding
rise at small $x$ in the structure functions too, once the off-shell
gluons have been appropriately coupled to the quark sector. This expectation
has been confirmed \cite{AKMS_mod,AKMS_rev} for a small range
of medium $Q^2$ even
in the case of running coupling BFKL, with $F_2\sim x^{-\lambda}$ and
$\lambda\gtrsim 0.5$ for $x<10^{-3}$.

Experimentally the situation has improved drastically since the
advent of the HERA $ep$ collider. Before HERA only the Fermilab
experiment E665 \cite{E665} was able to reach the small $x$ region,
but at rather
low $Q^2$. Now HERA takes data \cite{HERA_93,HERA_pre}
with $x/Q^2\gtrsim 3\cdot 10^{-5}$
over a wide range of $Q^2$. The observed strong rise of $F_2$
at small $x$ has boosted the interest in the BFKL formalism. On the
other hand the dynamically generated AP partons \cite{grv94} create a steep
gluon differently, via a long evolution length in $Q^2$, and successful
parameter-free predictions \cite{grv92,grv94,grv90} have been given long
before HERA started to operate. Alternatively, the present data can be
fitted using the NLO AP evolution equations: Then a term $\sim x^{-\lambda}$
has to be {\em assumed\/} for the gluon distribution,
e.\ g.\ in MRSG \cite{MRSG}. Possibly this
term mimicks the BFKL behaviour. But this is not clear, since both methods
describe the data equally well.

We conclude that a detailed comparison of the standard BFKL
formalism with the new data is necessary. Only this can tell us
if BFKL can rival conventional field theoretic renormalization group
(AP) evolution equations in describing the measured structure function
$F_2$.
Our calculations are based on the methods employed by Askew, Kwieci\'nski,
Martin and Sutton (AKMS) \cite{AKMS_mod,AKMS_rev,AKMS_box},
which will be described briefly in the following.

\subsection{\label{introsub} BFKL equation and $k_T$-factorization}
The unintegrated gluon distribution $f(x,k^2)$, which is related
to the familiar integrated gluon distribution used in the AP equations
by
\newcounter{temp}
\setcounter{temp}{\value{equation}}
\stepcounter{temp}
\def\theequation{\arabic{temp}\alph{equation}}
\setcounter{equation}{0}
\begin{eqnarray}
\label{fg_int}
x\, g(x,Q^2)&=&\int_0^{Q^2}\,\frac{dk^2}{k^2}\, f(x,k^2),\\
\label{fg_diff}
f(x,k^2)&=&\left.\frac{\partial x\, g(x,Q^2)}{\partial\ln(Q^2)}
\right|_{Q^2=k^2},
\end{eqnarray}
\setcounter{equation}{\value{temp}}
\def\theequation{\arabic{equation}}
depends on the transverse momentum $k$. Using it, one can write the BFKL
equation as \cite{nBFKL}
\begin{eqnarray*}
f(x,k^2)&=&f_0(x,k^2)+\int_x^1\,\frac{dy}{y}\,\int\,dk'^2\,
K(k^2,k'^2)\, f(y,k'),\\
K(k,k')&=&\frac{3\,\alpha_s(k^2)}{\pi}\, k^2\,\left\{\frac{1}{k'^2\,
|k^2-k'^2|}-\beta (k^2)\, \delta(k^2-k'^2)\right\},\\
\beta (k^2)&=&\int\,\frac{dk'^2}{k'^2}\, \left\{\frac{1}{|k^2-k'^2|}-\frac{1}{
(4\, k'^4+k^4)^\frac{1}{2}}\right\}.
\end{eqnarray*}
One could use a suitable input $f_0$, the so called ``driving term'',
and solve the equation iteratively \cite{KMS}, but this procedure
allows no simple connection to the known AP region.

Instead we can obtain an evolution equation in $x$ from the integral
equation by differentiating with respect to $\ln (1/x)$
\begin{equation}
\label{master}
-x\,\frac{\partial\, f(x,k^2)}{\partial x}=\frac{3\,\alpha_s (k^2)}{\pi}
\, k^2\int\,\frac{dk'^2}{k'^2}\,\left[\frac{f(x,k'^2)-f(x,k^2)}{|k'^2-k^2|}+
\frac{f(x,k^2)}{(4\, k'^4+k^4)^\frac{1}{2}}\right],
\end{equation}
assuming that the derivative of the driving term $f_0$
can be neglected. Since the driving term describes the gluon content
without any BFKL evolution, it is reasonable to assume that it is
connected to the non-perturbative ``soft'' pomeron \cite{SoftP}.
Because of the soft pomeron's weak $x$ dependence $\sim x^{-0.08}$, we
expect $\partial f_0/\partial\ln (1/x)$ to be small.

Eq.\ (\ref{master}) can then be used to evolve the unintegrated gluon
to smaller $x$,
using a suitably modified AP input as boundary condition at $x_0=10^{-2}$
by applying Eq.\  (\ref{fg_int}).
An obvious problem in Eq.\ (\ref{master}) is posed by the integration
over $k'^2$, which starts at zero. Even for the AP gluon distributions
we use in this work the limit of validity is about 1 \gev . We employ a simple
ansatz to continue the gluon into the infrared (IR) region,
which will be described later in this paper. A similar comment applies
to the upper limit of the $k'^2$ integration. The upper limit introduced
by energy conservation is close to infinity \cite{AKMS_mod,AKMS_rev},
so that for practical calculations an artificial ultraviolet cutoff
has to be introduced. The dependence on the IR and UV treatment will be
thoroughly discussed later on.

To obtain predictions for $F_2$, we have to convolute the (off-shell) BFKL
gluon,
i.\ e.\ the gluon ladder in Fig. \ref{leiter}, with the
photon-gluon fusion quark box $F^{(0)}$ using the $k_T$-factorization theorem
\cite{ColEl,Factor}
\begin{equation}
\label{kt}
F_i(x,Q^2)=\int\,\frac{dk'^2}{k'^4}\,\int_x^1\,\frac{dy}{y}\,
f\left(\frac{x}{y},k'^2\right)\, F_i^{(0)}(y,k'^2,Q^2),
\end{equation}
with $i=T,L$ denoting the transverse and longitudinal parts, respectively.
The expressions for $F^{(0)}$ can be found in \cite{AKMS_box}
and references therein. The BFKL prediction for $F_2$ is then simply
the sum of the calculated $F_T$ and $F_L$. Obviously here the same
problems with the $k'^2$ integration occur, which are circumvented by the
methods mentioned above.

It is important to notice that the BFKL gluons are {\em off-shell\/}
$(k^2\not=0)$. The term $F^{(0)}/k^2$ in the above equation then corresponds
to the structure function of a virtual gluon. In contrast the AP
formalism is based on {\em on-shell\/} gluons, which is a good approximation
due to the strong ordering in transverse momentum encountered in AP evolutions
in LO.
This strong ordering allows us to perform the $k'^2$ integration in
Eq.\ (\ref{kt}).
Thus, ignoring complications due to the collinear
singularities for simplicity, one arrives at the usual mass factorization
equation
\[ F_i(x,Q^2)=\int_x^1\,\frac{dy}{y}\, g\left(\frac{x}{y},Q^2\right)\,
\hat{F}_i(y,Q^2).\]
Here $\hat{F}_i$ plays the r\^{o}le of the on-shell gluon structure
function, whose $x$ dependence stems from the AP splitting function
$P_{qg}$, and $g$ is the integrated gluon.
In NLO AP evolutions the strong ordering in $k^2$ does not hold anymore due to
the emission of a second gluon, but the interacting gluon is still
considered to be on-shell in comparison with the hard scattering
scale $Q^2$.

This shows that it is inconsistent to simply feed the
evolved BFKL gluon via Eq.\  (\ref{fg_int}) back into the AP equations
below the limit set by $x_0$. Calculations attempting to use
BFKL gluons below and AP gluons above $x_0$ within the AP
formalism \cite{StroKot} ignore the essential {\em off-shellness\/}
of the BFKL gluons. This casts first doubts on a recent BFKL
analysis of the HERA H1 data using this method \cite{h1glue}.
No such problem persists when we use just this gluon mixture
to drive the general $k_T$-factorization Eq.\  (\ref{kt}), which
reduces to the mass factorization in the AP region.

Even though the dominant contribution at small $x$ should come from
the BFKL gluons, a certain amount of ``background'' in $F_2$ due to
quarks and non-perturbative effects should be taken into account.
We expect the background to be comparatively small and also to vary
much less with $x$. We shall use an ansatz motivated by the soft
pomeron $C_{I\! P}\, x^{-0.08}$, where the constant $C_{I\! P}$ is fitted to
the data.
After this general outline of our method, we will now proceed to a detailed
discussion of the underlying formalism.

\section{Suitable input for the BFKL evolution}

We focus our analysis on the gluon
distribution used in \cite{AKMS_rev}, i.e.\ a gluon
based on the MRS $D_0$-set of parton distributions \cite{d0d-},
but evolved with the leading order Altarelli-Parisi
equations \cite{Sutton}.
Since the BFKL evolution deals with an {\em unintegrated\/}
gluon distribution,
we calculate its derivative using the well-known singlet
Altarelli-Parisi equation given by
\[
f^{\mathrm{AP}}(x,Q^2)=\frac{\alpha_s(Q^2)}{2\pi}
	\int_x^1 \frac{dy}{y}\left(P^{(0)}_{gg}\left(\frac{x}{y}\right)\,
	g(y,Q^2)+
	P^{(0)}_{gq}\left(\frac{x}{y}\right)\,\Sigma(y,Q^2)\right),
\]
where $\Sigma$ denotes the quark singlet part, and $P^{(0)}_{gg}$,
$P^{(0)}_{gq}$
the usual LO splitting functions. In the same way we produce a leading
order MRS $D_-$-type gluon, based on an input given in \cite{d-type}.

We also use a dynamically generated gluon distribution, for definiteness
the GRV '92 LO parametrization \cite{grv92}, that has the
advantage of (a) being based on an explicit LO calculation and (b) being
positive definite down to a low value of $Q^2$.

Furthermore, we take a look at the MRSA-Low $Q^2$ gluon \cite{mrsalow}, which
extends the valid $Q^2$ range down to $0.625$~\gev\ using an ad
hoc form-factor-like
ansatz similar to the one we employ for $f$. However, as MRSA is a
NLO analysis, there is no consistent way to implement it in our BFKL evolution,
which is neither an $\overline{\mathrm{MS}}$- nor DIS-renormalization scheme
calculation.

\subsection{Treatment of the infrared region}

As mentioned in the Introduction, we need a suitable description of the
infrared region; for our calculations we use the ansatz explained in
detail in \cite{AKMS_mod,AKMS_rev}. This ansatz introduces three
parameters:
\begin{itemize}
  \item An IR-``cutoff'' $k_c^2$, i.e.\ a parameter which
   separates the infrared region, where an assumption on the $k^2$-behaviour
   of $f$ has to be made, from the region where the Lipatov equation
   is solved numerically.
  \item A parameter $k_a^2$ that controls the infrared behaviour of $f$.
   For $k^2 < k_c^2$ we set:
\begin{equation}
	f(x,k^2)= C' \frac{k^2}{k^2+k_a^2} f(x,k_c^2). \label{eqiransatz}
\end{equation}
   The proportionality constant is given by
\[
	C'=\frac{k_c^2+k_a^2}{k_c^2},
\]
   to guarantee continuity at $k^2=k_c^2$.
   This ansatz ensures that for $k^2\rightarrow 0$,
   $f(x,k^2)\sim k^2$, as required by gauge invariance.
  \item A scale $k_b^2$ where we ``freeze'' the running coupling constant,
   i.e.\
\[
	\alpha_s(k^2) \longrightarrow \alpha_s(k^2+k_b^2).
\]
\end{itemize}

This procedure applies also to our boundary
condition $f(x_0, k^2)$, although there is a slight modification to
the pure AP gluon in order to soften its low $k^2$ behaviour:
\begin{equation}
	f^{\mathrm{AP}}(x_0,k^2) \longrightarrow f^{\mathrm{AP}}
	(x_0,k^2+k_s^2) \label{eqshift}.
\end{equation}

The only purpose of the additional parameter $k_s^2$
is to ensure that we do not approach too closely a region where the gluon is
unreliable; e.g.\ the $D_0$-type gluon already approaches zero at
$Q^2=1$~\gev, and the $D_-$-type gluon is not defined at all for such a low
scale.

Thus,
\begin{equation}
  f(x_0,k^2) = \left\{ \begin{array}{ll}
		   	\left(\frac{k^2}{k^2+k_a^2}\right)f^{\mathrm{AP}}
			(x_0,k_c^2+k_s^2) & \quad\mbox{for~} k^2 < k_c^2 \\
			\left(\frac{k^2}{k^2+k_a^2}\right)f^{\mathrm{AP}}
			(x_0,k^2+k_s^2) & \quad\mbox{for~} k^2 \ge k_c^2. \\
			\end{array} \right. \label{eqbound}
\end{equation}
This provides an infrared behaviour according to (\ref{eqiransatz}), and for
$k^2$ large enough $f(x_0,k^2)$ approaches $f^{\mathrm{AP}}(x_0,k^2)$.

In ref.\ \cite{AKMS_mod}, the parameter $k_s^2$ is set equal to
$k_a^2$, whereas in \cite{AKMS_rev} $k_s^2$ equals $k_b^2$. We prefer setting
$k_s^2=k_a^2$ for reasons given below.

\subsection{Fixing $k_a^2$}

It seems to be clear that the parameters introduced above should be small;
so the simplest choice for these would be \cite{AKMS_rev}:
\begin{equation}
	k_a^2 =  k_b^2 =  k_c^2 = k_s^2 = 1\mbox{~GeV}^2. \label{eqsimple}
\end{equation}

However, there is a self-consistency constraint on the choice of $k_a^2$,
depending on
$k_c^2$ and $x_0$. The boundary condition (\ref{eqbound}) should inversely be
related
to the (integrated) Altarelli-Parisi gluon $g$ in Eq.\ (\ref{fg_int}), that is
\begin{equation}
  x_0\,g(x_0,Q^2)-\int_0^{Q^2} \frac{d k^2}{k^2} f(x_0,k^2)=0. \label{eqconst}
\end{equation}
In a stricter approach, we apply the shift introduced in (\ref{eqshift}) also
to the integrated gluon $g$ in this equation. Then it is not very difficult to
show that
\begin{equation}
\int_{k_c^2+k_a^2}^{Q^2+k_a^2}dl^2 \left(1+\frac{k_s^2-k_a^2}{l^2}\right)
\left.\frac{\partial x_0\,g(x_0,Q^2)}{\partial Q^2}\right|_{Q^2=
l^2+k_s^2-k_a^2}=x_0\,g(x_0,Q^2+k_s^2)-x_0\,g(x_0,k_c^2+k_s^2).
\label{eqdetks}
\end{equation}
It is obvious that this equation is fulfilled by setting $k_s^2=k_a^2$, hence
motivating our previous assumption. Using this and the asymptotic behaviour
of $g$,
\begin{equation}
  x\,g(x,Q^2+k_s^2) \stackrel{Q^2\gg k_s^2}{\simeq} x\,g(x,Q^2),
\end{equation}
we may return to (\ref{eqconst}) for simplicity, which gives us upon
inserting our boundary condition (\ref{eqbound})
\begin{eqnarray*}
  x_0\,g(x_0,Q^2) & = &
\int_0^{k_c^2} \frac{d k^2}{k^2+k_a^2} f^{\mathrm{AP}}(x_0,k_c^2+k_a^2)
+ \int_{k_c^2}^{Q^2} \frac{d k^2}{k^2+k_a^2} f^{\mathrm{AP}}(x_0,k^2+k_a^2) \\
 & \simeq & f^{\mathrm{AP}}(x_0,k_c^2+k_a^2)\ln\frac{k_c^2+k_a^2}{k_a^2}+
       x_0\,g(x_0,Q^2)-x_0\,g(x_0,k_c^2+k_a^2).
\end{eqnarray*}
Thus we obtain an implicit equation for $k_a^2$:
\begin{equation}
\ln\frac{k_c^2+k_a^2}{k_a^2}\simeq\frac{x_0\,g(x_0,k_c^2+k_a^2)}
{f^{\mathrm{AP}}(x_0,k_c^2+k_a^2)}. \label{eqestka}
\end{equation}
One can see that the value of $k_a^2$ mainly depends
on the ratio $g/f$ in the region of low $Q^2$, and as the variation of $f$ for
different parton distributions is relatively small in that region compared to
the variation of $g$, we conclude that it is basically the absolute value of
$g$ that determines the size of $k_a^2$.
Hence, as a rule of thumb, the higher
the value of $x_0\,g(x_0,Q^2)$ for small $Q^2$, the smaller the resulting
$k_a^2$. It should be emphasized that, although the shift in Eq.\
(\ref{eqshift})
is taken into account in Eq.\ (\ref{eqdetks}) for both $f$ and $g$, but in
Eq.\ (\ref{eqestka}) only for $f$, the
solution of Eq. (\ref{eqestka}) provides a good estimate on $k_a^2$,
very close to the value we get by minimizing the l.h.s.\ of Eq.\
(\ref{eqconst}) for $Q^2>100$~\gev.
The result for all the parton distributions under consideration is given in
Table \ref{tbparam}, together with the value of
\lqcd\ for four flavors used with each distribution.
Figure \ref{bound} shows the unmodified gluons and the modified boundary
conditions according to Table \ref{tbparam}.

Regarding the $D_0$-type gluon, we see that the optimized value of
$k_a^2=0.95$~\gev\ is indeed very close to the na\"\i ve estimate of 1~\gev, as
was already noticed in \cite{AKMS_rev}.
It was also mentioned there that a reasonable choice of
$k_a^2$ should lie in the
range of $0.5 - 2$~\gev, and we see that the $D_-$-type value lies well
within this range,
while the MRSA set is already close to its lower edge. The most extreme
boundary condition in this respect is derived from the GRV parametrization,
with a value of only $0.19$~\gev.

\begin{table}
	\centering
	\begin{tabular}{|c|c|c|c|c|}
	\hline
	\oustrut Set & $\Lambda_{\mathrm{QCD}}^{(4)}$~[MeV] & $x_0$ &
        $k_c^2$~[\gev] & $k_a^2$~[\gev] \\
	\hline
	\hline
	\oustrut $D_0$-type & 173.2 & 0.01 & 1.0 & 0.95 \\
	\oustrut $D_-$-type & 230.4 & 0.01 & 1.0 & 1.51 \\
	\oustrut GRV '92-LO & 200.0 & 0.01 & 1.0 & 0.19 \\
	\oustrut MRSA-Low $Q^2$ & 230.0 ($\overline{\mathrm{MS}})$ & 0.01 &
        1.0 & 0.44 \\
	\hline
	\end{tabular}
	\caption{Parameters for various sets of parton distributions}
	\label{tbparam}
\end{table}

As the more recent parton distributions favor a larger gluon $g$, the
assumption of $k_a^2=1$~\gev\ (a good choice for the relatively small
$D_0$-type gluon) does not appear to be the best choice for these gluons.

Let us now emphasize the importance of the constraint (\ref{eqconst}) on
$k_a^2$: Starting with the simple assumption (\ref{eqsimple}) that all the
parameters introduced should be equal to 1~\gev, one gets a slope
\[
	\lambda = \left.\frac{1}{f}\frac{\partial f}{\partial\ln(1/x)}
	\right|_{x=10^{-4}} = 0.5 - 0.6
\]
for the unintegrated, BFKL evolved gluon distribution,
as expected, no matter if $D_0$-type, $D_-$-type, or GRV is chosen
as input for the BFKL evolution. If we deviate each of the infrared parameters
from (\ref{eqsimple}), we see that it is $k_a^2$ which has the biggest impact
on $\lambda$. With $k_a^2$ decreasing, the slope rises, resulting in a slope
$\lambda\simeq 0.9$ for GRV partons ($k_a^2=0.19$~\gev) as the extreme limit.

Keeping these considerations in mind, we will now concentrate our analysis on
the structure functions. We will demonstrate that the slopes of $f$ and
$F_2$ are related in such a way that for large $\lambda$ our calculated
$F_2$ is too steep to match it to the recent HERA data.

\section{Varying the parameters}
We construct the boundary condition at $x_0=10^{-2}$.
The consistency constraint (\ref{eqconst}) determines $k_a^2$ and
as a standard value for the other two IR parameters we choose $k_b^2=k_c^2=
1\,\mbox{GeV}^2$. The UV cutoff is set to $10^4\, \mbox{GeV}^2$
and we stay with the \lqcd\ of the AP partons. We have written an evolution
program to solve (\ref{master}) iteratively.
Below $k_c^2$ the necessary
integration can easily be done analytically, above we use
the Gau{\ss}-Legendre quadrature. This transforms the integro-differential
equation (\ref{master}) into a set of coupled differential equations,
allowing us to use the standard Runge-Kutta method to calculate the
evolution. The evolved gluon below $x_0$ is combined with the unintegrated
AP gluon above $x_0$ to obtain predictions
for $F_T$ and $F_L$ by performing the integration in Eq.\ (\ref{kt}) with
Monte-Carlo methods. For
convenience a fit of $F_2=F_T+F_L$ is given
in the Appendix. Finally we fit the background to the data, as discussed at
the end of Section \ref{introsub}, and obtain our BFKL prediction.

It is vital to check the dependence of the results on the parameters
used for the IR and UV treatment. In Fig. \ref{vary} we varied all relevant
parameters, using the $D_0$-type gluon. All curves have been calculated
at $Q^2=15\,\mbox{GeV}^2$ and are already fitted to the shown HERA
data with the soft pomeron background $C_{I\! P}\, x^{-0.08}$.

The strong dependence on $k_a^2$, which we expect from the
corresponding variation in the gluon slope, is obvious. The curves for
low $k_a^2$ are much too steep. At $k_a^2=0.2$ \gev\
for example, the pure BFKL prediction is too high, even without any background.
A background fit would then give a {\em negative\/} contribution, i.\ e.\
$C_{I\! P}<0$,
which is unphysical. In all such cases we set the background to zero.
It is crucial, that $k_a^2$ can be precisely determined from the
consistency constraint. Without the constraint, we could vary the slope
of $F_2$ from 0.5 to 0.8 by choosing $k_a^2$ within the shown range
of 0.2 to 2.0 \gev ,
rendering any serious prediction impossible. In the already mentioned
comparison of a BFKL calculation with HERA data \cite{h1glue}, $k_a^2$
is treated as a {\em free\/} parameter. If the fitted $k_a^2$ should not
be close to the consistent value by chance, it is doubtful that
any strong conclusions can be drawn from a successful description of the data.

The influence of the IR cutoff $k_c^2$ is comparably small. It should
be kept in mind that $k_a^2$ has to be fitted separately for each $k_c^2$.
Actually the induced variations in $k_a^2$ dampen the dependence
of the predictions on $k_c^2$ slightly. It is no surprise, that
variations of the UV cutoff do not introduce much uncertainty into the
predictions, as the running coupling already serves as an effective
UV cutoff \cite{AKMS_mod,AKMS_rev}. The remaining free IR parameter,
$k_b^2$, has a sizeable effect on the curves. The slope of $F_2$ varies from
0.53 to 0.58 in the shown $Q^2$ range. The standard value of 1 \gev\
gives a slope of about 0.55, and the variations of $k_b^2$ represent an
effective error-band of our calculations.

A considerable dependence on \lqcd\ is expected and can be
seen in Fig. \ref{vary}. Fortunately this parameter is fixed, since we are
using AP gluons with a given \lqcd\ as boundary condition. It is
interesting to note, that higher values of \lqcd\ than the rather
low 173.2 MeV used in the $D_0$-type partons give {\em un\/}favourably steep
slopes! Finally we take a look at variations of $x_0$, noting that
we have to determine $k_a^2$ separately again.
We expect the steep curve for $x_0=5\cdot 10^{-2}$ due to the long
evolution length in $x$. But it is daring to use the BFKL equations at
such high values of $x$ and the data do not support such a choice.
We can also see that the comparably flat curve created
by a short evolution in $x$ starting from $x_0=10^{-3}$ does fit the data
well. But this success can obviously not be claimed by the BFKL evolution,
since the data do not extend far below that $x_0$. In order to test if
the BFKL equations can describe the {\em current\/} data, we have to choose
a larger $x_0\sim 10^{-2}$. The typical steep BFKL slope simply needs enough
evolution length to develop. Small variations
to lower $x_0$ from the usually used $10^{-2}$ do not affect the
calculations strongly. The standard values \cite{AKMS_mod,AKMS_rev} used for
$k_b^2$ and $x_0$ give slopes slightly below and above the
average expected from the variations, respectively.
Thus this choice of parameters
will give a sensible prediction while still allowing comparisons with
earlier calculations \cite{AKMS_mod,AKMS_rev}.

The results presented in this sections show that the dependence
of the calculations on the parameters is under control. This
statement would be impossible, if $k_a^2$ did not obey the
consistency constraint (\ref{eqconst}).

\subsection{Using other input distributions}

Besides our detailed analysis of the $D_0$-type gluon, we also studied
the effect of feeding different parton distributions into our evolution,
namely GRV'92 LO \cite{grv92} and MRSA-Low~$Q^2$ \cite{mrsalow}. The results
for $F_2$ can be seen in Fig.\ \ref{mrsa}.

Obviously by using GRV and MRSA gluons as inputs for the BFKL evolution, a
structure function is generated which is far too steep
for the data. Even though we have got, in principle, some freedom in fitting
an appropriate background, this is useless here, since already the pure
BFKL part of $F_2$ is too high for the HERA data, so that we must set
the background to zero.

The steepness of $F_2$ is closely related to the fact that the proper
$k_a^2$ is much smaller than 1~\gev, resulting in a larger
slope $\lambda$ of the BFKL gluon $f$. As we explained above, this is mainly
an effect of the size of the integrated gluon input $g$. This suggests the
conclusion that the
whole BFKL procedure has only a chance to work with the older (smaller) gluons.
Modern parton distributions imply small values of
the crucial infrared parameter $k_a^2$,
and since this has a strong effect on the calculated structure functions,
it is likely that one does not succeed in matching these to present
experimental data.

\section{Comparison with data}
In Fig. \ref{e6pre} we compare our calculations with very low $Q^2$ data from
the Fermilab E665 experiment \cite{E665}.
Preliminary data of the HERA $ep$ collider \cite{HERA_pre} at low $Q^2$
are also shown in this figure. The published 1993 HERA data \cite{HERA_93}
for low to medium $Q^2$ are presented in Fig. \ref{hera} together with our BFKL
predictions. All graphs show the BFKL calculations based on the $D_0$-type
and the $D_-$-type partons, as well as the latest dynamical NLO renormalization
group predictions (GRV), as
presented in \cite{grv94}. Also shown is the soft pomeron background,
which is included in the $D_0$-type curve. A general feature of all
figures is that the difference between the $D_0$-type and the $D_-$-type
BFKL predictions is very small after fitting the background.
For this reason we do not discuss them separately. We have
checked, that the $R=F_L/F_T$ values used to extract $F_2$ from
the experimental data are close enough to those predicted by BFKL.
Thus it is not necessary to
reanalyze the data in terms of $R_{\mathrm{BFKL}}$.

We first turn our attention to the E665 data \cite{E665}.
The data do not extend very far into the small $x$ region, making
a check of the BFKL behaviour difficult. On the other hand the difference
between the GRV (AP) and BFKL predictions is potentially large at small $x$.
We also notice that the added background is comparable in size to the
BFKL part at higher $x$. This flattens the steep BFKL behaviour, giving
a good description of the data. But it is just the large contribution
of the background which makes the very procedure used doubtful.

A further hint, that the successful description of the E665 data should
not be taken too seriously, is provided by comparing the curves for
the E665 data at 2.8 \gev\  and those for preliminary ZEUS
data at 3.0 \gev . While the E665 background is strong $(C_{I\! P}=0.189)$
for $D_0$-type, the optimal ZEUS background would be negative
$(C_{I\! P}=-0.172)$
and therefore is set to zero. Thus the natural requirement, that the BFKL
predictions for similar $Q^2$ values should be approximately equal,
is only fulfilled at very small $x$. For $x\gtrsim 10^{-4}$ the
influence of the background quickly becomes stronger and the curves deviate.
We conclude that BFKL can {\em not\/} describe both data sets consistently.
It is also obvious from the figures that just this is possible using the
usual NLO renormalization group equations, see the curves labeled GRV '94
in Fig.\ \ref{e6pre}.

Turning to the preliminary HERA data in Fig. \ref{e6pre},
we find that the BFKL slope of $F_2$ is evidently too steep. The same
tendency can also be found in the 1993 HERA data up to approximately
15 \gev\ as shown in Fig. \ref{hera}. Everywhere
in this region the background is very small or would even be negative,
if it were allowed. Due to the larger spread in $x$, the preliminary
data show the discrepancy rather clearly. The problem is rooted in
the almost $Q^2$ independent slope of BFKL, which is always $\gtrsim 0.5$
and grows weakly with $Q^2$. The dynamical GRV (AP) calculations give on
the contrary
flat slopes at low $Q^2$, which rise significantly with larger $Q^2$.
This is obviously in much better agreement with the data.
For example, using the preliminary ZEUS data at 4.5 \gev ,
we get a total $\chi^2$ of 15.3 ($D_0$-type), 15.0 ($D_-$-type)
and of 2.81 (GRV '94 NLO) for the four data points.

At $Q^2$ higher than 15 \gev , we see that
the slope of the data becomes compatible with the one predicted
by BFKL. The rising slope of the dynamical GRV prediction fits the data
at least as well. It is interesting to note that even at higher
$Q^2$ a slight extension in the $x$ range could provide an indication
which evolution should be used. If future data should conform to the
already visible tendency that a Lipatov-like slope is only obtainable at
high $Q^2$, say $\gtrsim 50$ \gev , then the simple LO BFKL
formalism would become implausible, since its validity at such high $Q^2$ is
questionable.

\section{Conclusions}
We have shown that the AKMS method for calculating BFKL predictions
of $F_2$ remains stable under variations of the introduced parameters,
if the consistency constraint Eq.\ (\ref{eqconst}) on $k_a^2$ is applied.
As boundary
conditions one has to choose older, i.\ e.\ smaller, gluon densities, since
the large recent AP gluons lead to small $k_a^2$, which in turn produces
overly steep slopes of $F_2$. The BFKL boost at small $x$ is simply
too large for gluons constructed to produce the measured large $F_2$
slope via the conventional renormalization group (AP) evolution equations.

Thus we use the old $D_0$-type and $D_-$-type gluons with
consistently fixed $k_a^2$ as input for the BFKL evolution.
A further complication is introduced by
the necessity to add a background contribution to the BFKL prediction for
$F_2$. Then it seems
reasonable to require that the main growth of $F_2$ is not driven
by the chosen background, and that this background is comparably small. The
last condition is not fulfilled in the region of the E665 experiment,
casting serious doubts on the good agreement with the data.

In the HERA region we find in contrast small background contributions,
allowing for
reliable comparisons with experiment. Especially the preliminary HERA
data in Fig. \ref{e6pre} show that
the almost $Q^2$ independent BFKL slope is too steep for $Q^2\lesssim 15$
\gev . But even up to the expected limit of applicability in $Q^2$ of
the BFKL evolution we find that the predicted slope is somewhat too steep.
With improved statistics and maybe a slightly extended coverage
in $x$, HERA should be able to assess the LO BFKL predictions for
$F_2$. If the tendency visible in the current data is an
indication for future developments, we expect LO BFKL to fail the test.
It remains to be seen whether extensions of LO BFKL --- inclusion of the
quark sector, NLO BFKL or theoretically consistent (energy-momentum
conservation, etc.\ ) unified evolution equations --- can
improve the agreement with the data.
Our results also indicate that conventional (dynamical) renormalization group
evolutions are still the best method for calculating and analyzing $F_2$.

\section{Acknowledgments}
We are grateful to E.\ Reya and M.\ Gl\"uck for helpful guidance
and fruitful discussions, and to E.\ Reya, M.\ Stratmann and S.\ Kretzer
for carefully reading the manuscript. We thank P.\ J.\ Sutton for making
the $D_0$-type partons available to us. I.\ B.\ wishes
to thank D.\ Wegener for unbureaucratically granting additional
computing time on his workstations.
This work has been supported in part by the 'Bundesministerium f\"ur Bildung,
Wissenschaft, Forschung und Technologie', Bonn.

\section*{Appendix: Parametrization}
\setcounter{equation}{0}
\def\theequation{A.\arabic{equation}}
It should be convenient to have a simple parametrization of our theoretical
results for $F_2$.
The following is a parametrization of the BFKL part only,
to which an appropriate background still has to be added.

We use an ansatz of the form:
\begin{equation}
\label{ansatz}
F_2^{BFKL}=\alpha\, x^{-\lambda}+\beta\, x+\gamma,
\end{equation}
which describes all curves shown in Figs. \ref{vary} -- \ref{hera} well.
It is even
possible to parametrize {\em all\/} pure BFKL results for the $D_0$-type
and $D_-$-type gluons shown in Figs. \ref{e6pre} and \ref{hera} by
choosing the following $Q^2$ dependence
of the coefficients [$t=\ln (Q^2/\Lambda_{\mathrm{QCD}}^2)$]
\begin{eqnarray}
\label{anskoeff}
\alpha (Q^2)&=&A+B\, t+C\, t^2+D\, t^3,\nonumber\\
\beta (Q^2)&=&E\, t^2+F\, t^3,\\
\gamma (Q^2)&=&G+H\, (Q^2/\mathrm{GeV}^2) +I\, t,\nonumber\\
\lambda (Q^2)&=&J+K\, t.\nonumber
\end{eqnarray}
The corresponding coefficients are given in Table \ref{koeff}.
It is interesting to note that a small growth of $\lambda$ with $Q^2$
has to be taken into account.

\begin{table}
\begin{center}
\begin{tabular}{|c||r|r|}
\hline
\oustrut&\multicolumn{1}{c|}{$D_0$-type}&\multicolumn{1}{c|}{$D_-$-type}\\
\hline\hline
$\oustrut A$&$4.818\cdot 10^{-3}$&$8.07\cdot 10^{-4}$\\ \hline
$\oustrut B$&$-2.460\cdot 10^{-3}$&$4.1\cdot 10^{-5}$\\ \hline
$\oustrut C$&$7.386\cdot 10^{-4}$&$1.579\cdot 10^{-4}$\\ \hline
$\oustrut D$&$3.32\cdot 10^{-6}$&$4.370\cdot 10^{-5}$\\ \hline
$\oustrut E$&$-2.23\cdot 10^{-1}$&$-1.89\cdot 10^{-1}$\\ \hline
$\oustrut F$&$1.15\cdot 10^{-2}$&$1.86\cdot 10^{-2}$\\ \hline
$\oustrut G$&$-5.3\cdot 10^{-3}$&$-1.46\cdot 10^{-2}$\\ \hline
$\oustrut H$&$-6.66\cdot 10^{-4}$&$-8.72\cdot 10^{-4}$\\ \hline
$\oustrut I$&$6.54\cdot 10^{-3}$&$1.026\cdot 10^{-2}$\\ \hline
$\oustrut J$&$-5.113\cdot 10^{-1}$&$-5.700\cdot 10^{-1}$\\ \hline
$\oustrut K$&$-7.10\cdot 10^{-3}$&$-2.89\cdot 10^{-3}$\\ \hline
\oustrut \lqcd & 173.2 MeV & 230.4 MeV \\ \hline
\end{tabular}
\caption{\label{koeff} Coefficients defined in (\ref{anskoeff}) of the
parametrization (\ref{ansatz})}
\end{center}
\end{table}

It has to be stressed, that this parametrization is {\em only\/}
valid within the range $10^{-5}\le x\le 10^{-2}$ and 0.8 \gev\
$\le Q^2\le$ 120 \gev . On the one hand BFKL is not expected to be applicable
even at the edges of this region. On the other hand we note, that the
form of the ansatz has been tailored for this region only. The term
$\beta\, x$, for example, will lead to wrong results for $x>10^{-2}$.
%
%
%

\newpage
%
%
%
\section*{Figure Captions}
\begin{figure}
\begin{center}
\epsfig{file=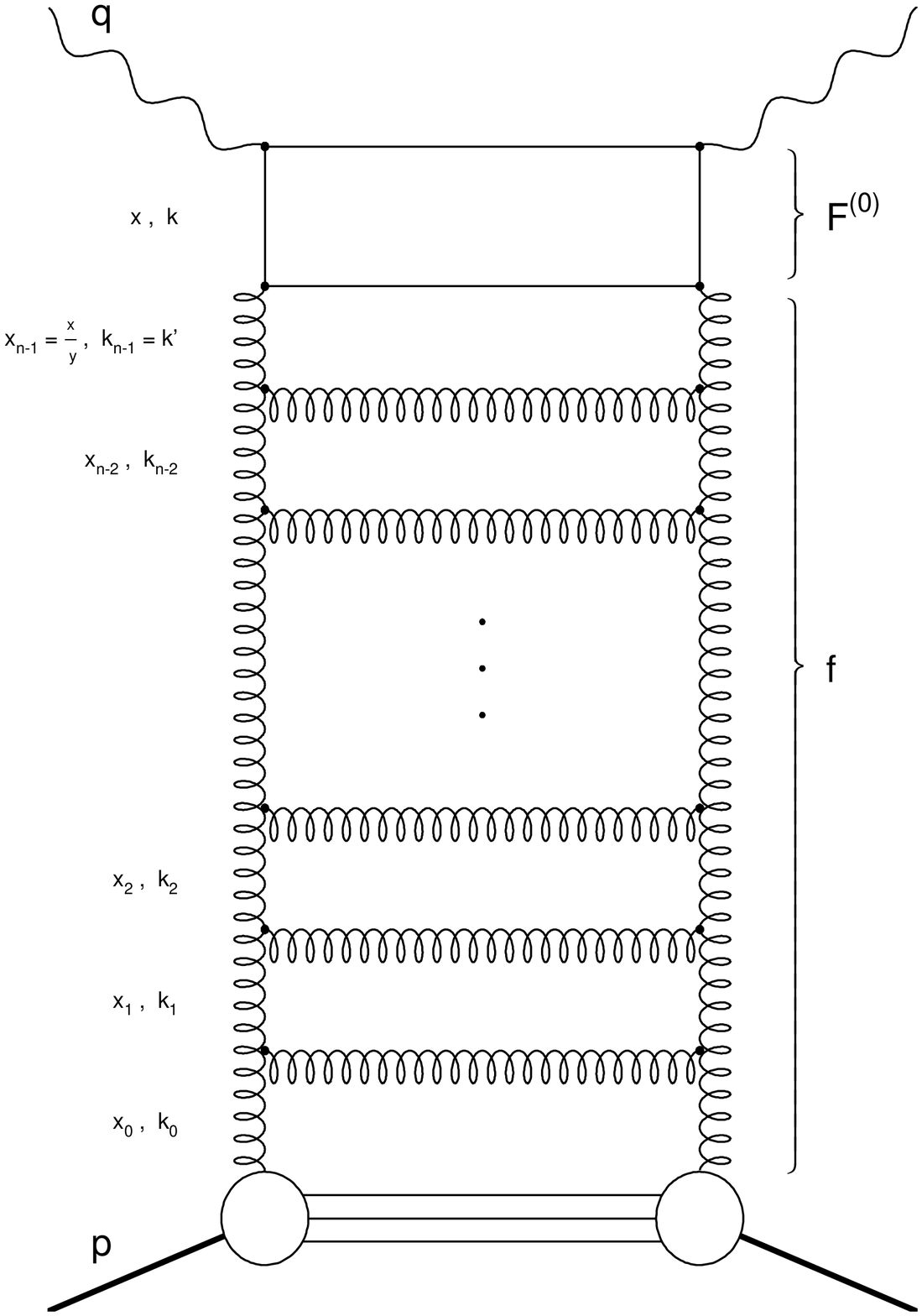,height=20.35cm,width=16cm}\\
\refstepcounter{figure}
\label{leiter}
\vspace{0.5cm}
{\large\bf Fig.\ \thefigure}
\end{center}
\end{figure}

\begin{figure}
\begin{center}
\epsfig{file=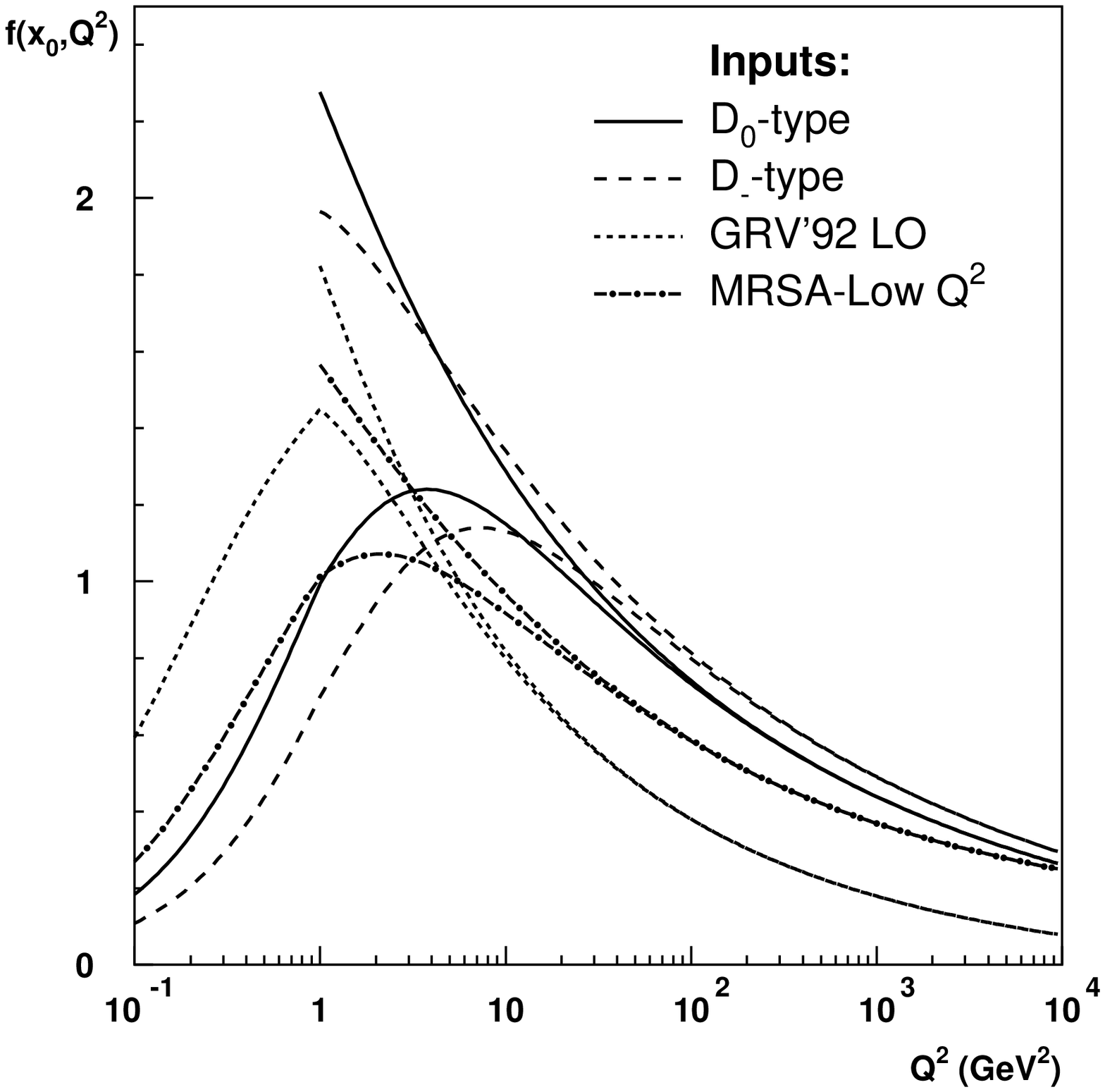,height=16cm,width=16cm}\\
\refstepcounter{figure}
\label{bound}
\vspace{0.5cm}
{\large\bf Fig.\ \thefigure}
\end{center}
\end{figure}

\begin{figure}
\begin{center}
\epsfig{file=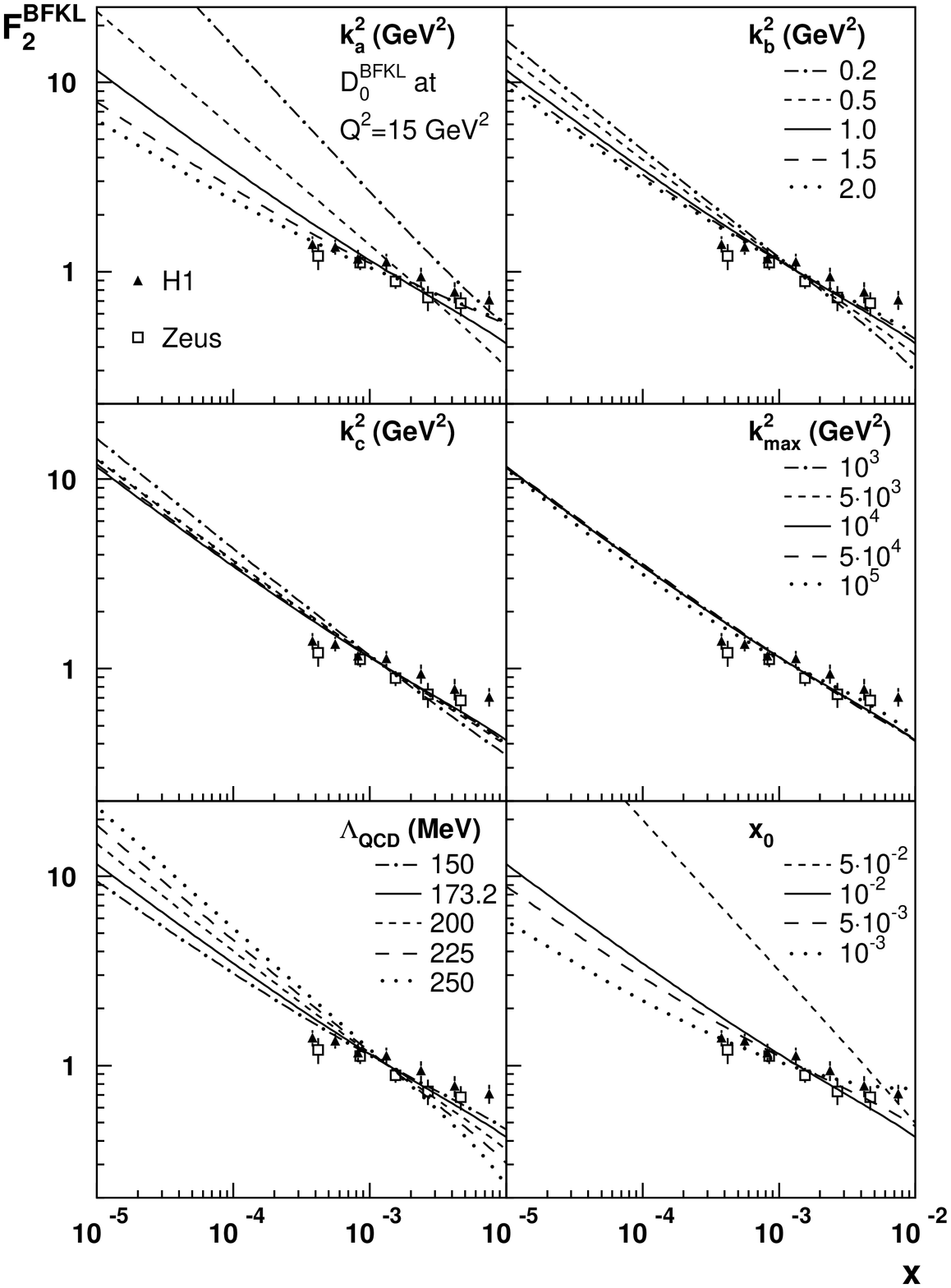,height=20.35cm,width=16cm}\\
\refstepcounter{figure}
\label{vary}
\vspace{0.5cm}
{\large\bf Fig.\ \thefigure}
\end{center}
\end{figure}

\begin{figure}
\begin{center}
\epsfig{file=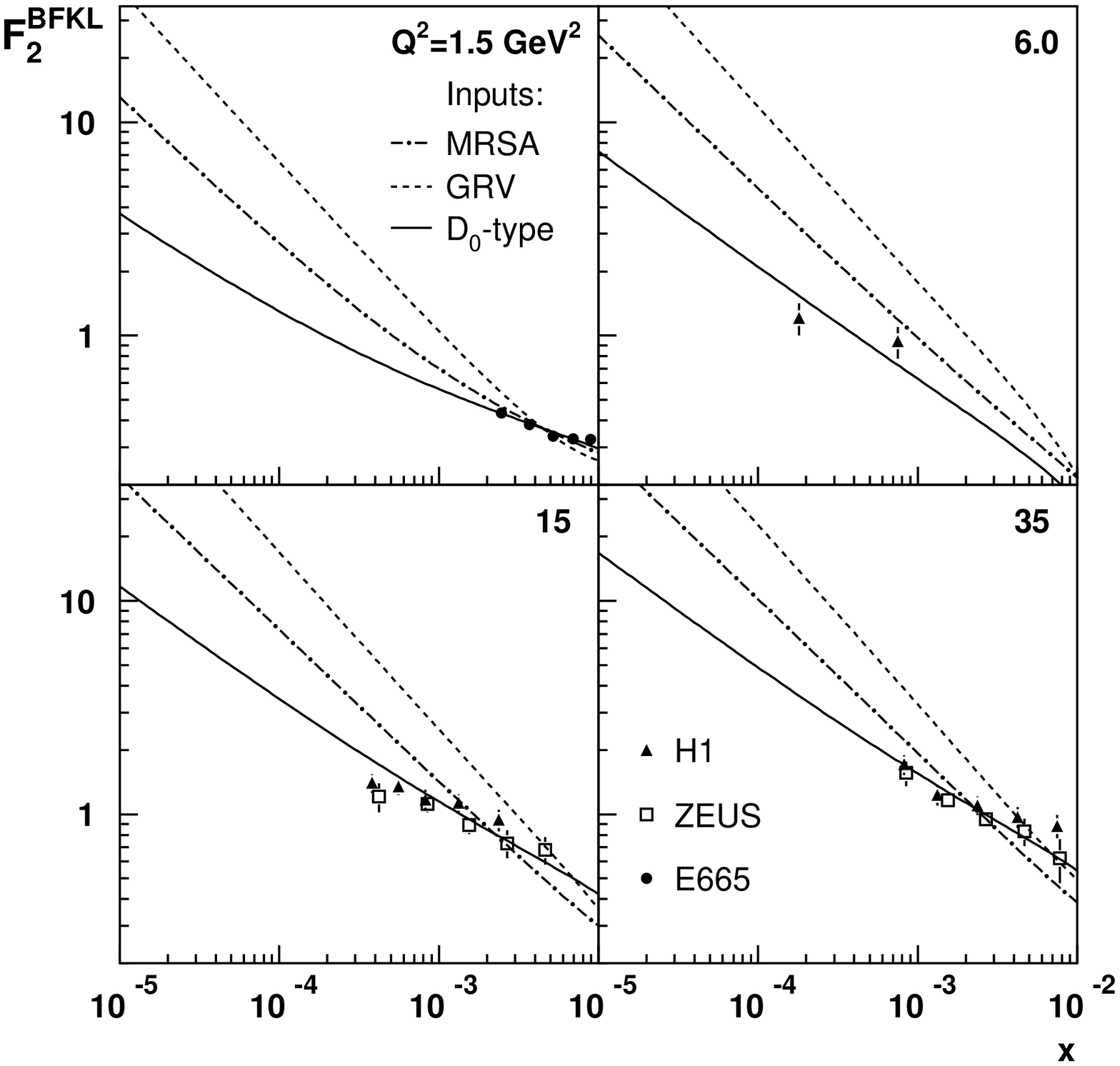,height=16cm,width=16cm}\\
\refstepcounter{figure}
\label{mrsa}
\vspace{0.5cm}
{\large\bf Fig. \thefigure}
\end{center}
\end{figure}

\begin{figure}
\begin{center}
\epsfig{file=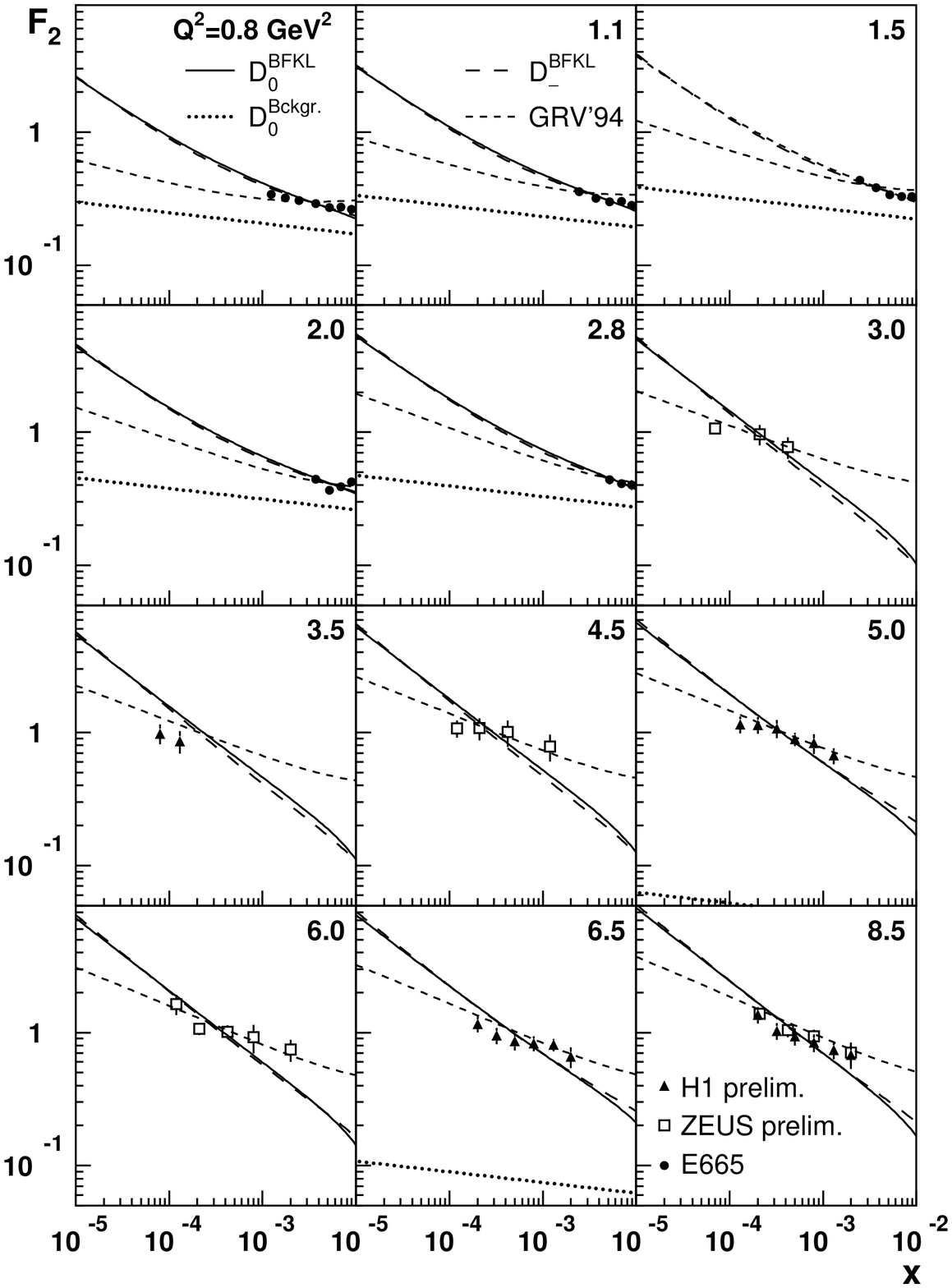,height=20.35cm,width=16cm}\\
\refstepcounter{figure}
\label{e6pre}
\vspace{0.5cm}
{\large\bf Fig.\ \thefigure}
\end{center}
\end{figure}

\begin{figure}
\begin{center}
\epsfig{file=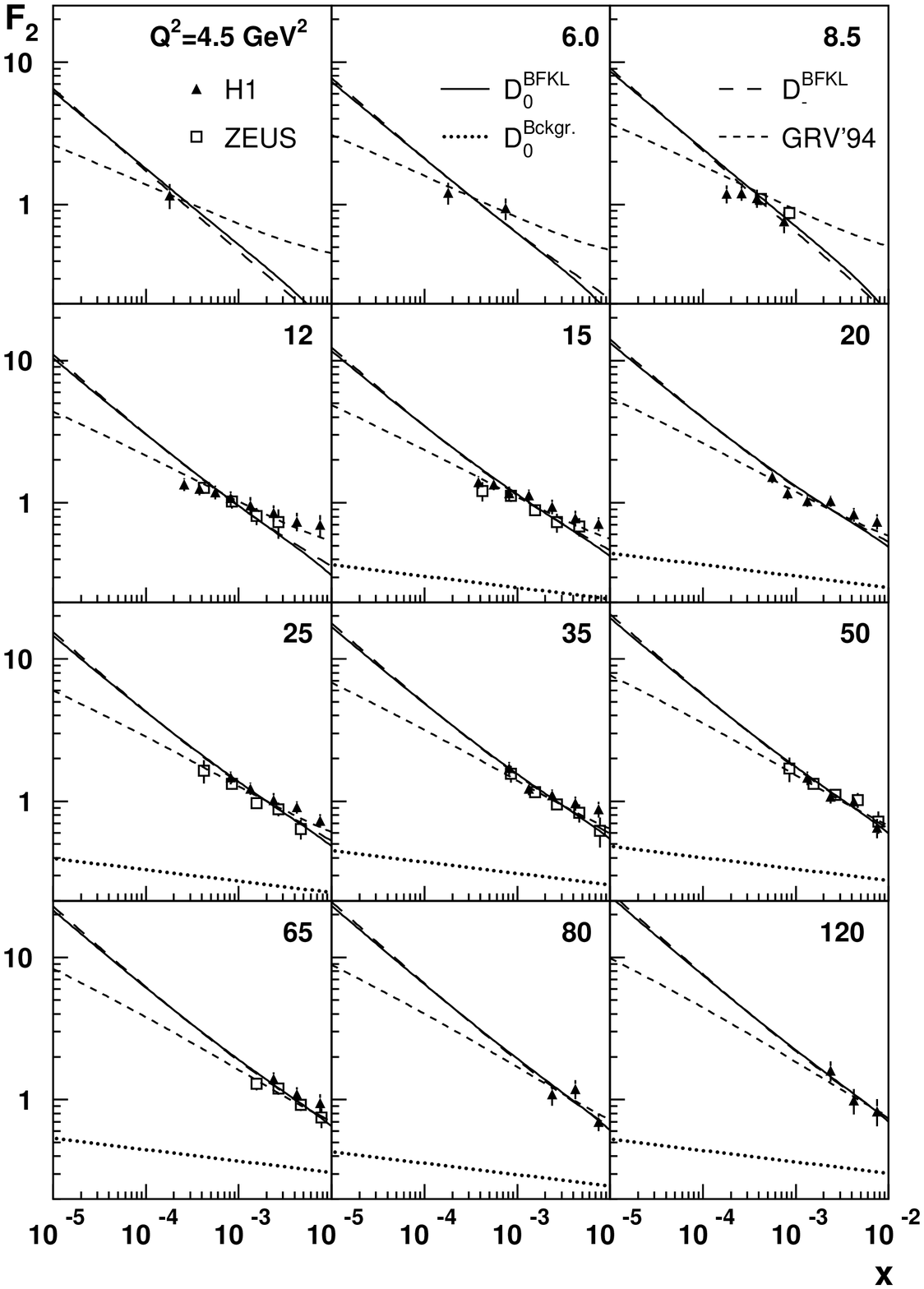,height=20.35cm,width=16cm}\\
\refstepcounter{figure}
\label{hera}
\vspace{0.5cm}
{\large\bf Fig.\ \thefigure}
\end{center}
\end{figure}

\begin{description}
\item[Fig.\ \ref{leiter}]
Gluon ladder with the quark box for deep-inelastic $ep$ scattering attached.
The momenta are shown on the left side of the Feynman diagram and the
parts of the $k_T$-factorization theorem, Eq.\ (\ref{kt}), on the right.
\item[Fig.\ \ref{bound}]
AP inputs as indicated and the corresponding boundary conditions constructed
via Eq.\ (\ref{eqbound}).
The unmodified AP
gluons are shown down to $k_c^2=1$ \gev, whereas the
IR treated inputs are continued below $k_c^2$.
\item[Fig.\ \ref{vary}]
Variations of the parameters. All curves are calculated at $Q^2=15$ \gev\ with
the $D_0$-type gluon input.
The background $C_{I\! P}\, x^{-0.08}$ has already been added to the BFKL
curves and statistical
and systematic errors of the data have been added in quadrature in this
and all following figures. The legend for $k_b^2$ is the same also for $k_a^2$
and $k_c^2$.
\item[Fig.\ \ref{mrsa}]
Comparison of BFKL results, using MRSA-Low $Q^2$ and GRV '92 LO gluon inputs,
compared with data at several $Q^2$. The
corresponding curves for the $D_0$-type gluon input is shown for comparison.
\item[Fig.\ \ref{e6pre}]
Low $Q^2$ data of the Fermilab E665 experiment \cite{E665}
and preliminary data of the HERA experiments \cite{HERA_pre} in
comparison with our BFKL results. The background
included in the $D_0$-type curves is separately displayed.
Note that GRV '94 refers to the conventional dynamical results
based on NLO AP evolutions of valence-like input parton densities \cite{grv94}.
\item[Fig.\ \ref{hera}]
As in Fig.\ \ref{e6pre} but using the HERA data of ref.\ \cite{HERA_93}.
\end{description}
%
%
%
\pagestyle{empty}
\vspace*{\fill}

\vspace*{\fill}

\vspace*{\fill}

\vspace*{\fill}

\vspace*{\fill}

\vspace*{\fill}
\end{document}